\documentclass[runningheads]{llncs}
\usepackage[T1]{fontenc}
\usepackage{caption}
\usepackage{subcaption}

\usepackage{graphicx}
\usepackage{pgfplots}
\pgfplotsset{width=10cm,compat=1.9}
\usepackage{float}
\usepackage{array}
\newcolumntype{H}{>{\setbox0=\hbox\bgroup}c<{\egroup}@{}}

\usepackage{cite}
\usepackage{amsmath,amssymb,amsfonts}
\usepackage{algorithmic}
\usepackage{graphicx}
\usepackage{textcomp}
\usepackage{xcolor}
\usepackage{listings}

\usepackage{booktabs}
\usepackage{multirow}
\usepackage{siunitx}

\definecolor{codegreen}{rgb}{0,0.6,0}
\definecolor{codegray}{rgb}{0.5,0.5,0.5}
\definecolor{codepurple}{rgb}{0.58,0,0.82}
\definecolor{backcolour}{rgb}{0.95,0.95,0.95}

\lstdefinestyle{mystyle}{
    backgroundcolor=\color{backcolour},   
    commentstyle=\color{codegreen},
    keywordstyle=\color{magenta},
    keywordstyle=[2]\color{blue},
    numberstyle=\tiny\color{codegray},
    stringstyle=\color{codepurple},
    basicstyle=\small\ttfamily,
    language=C,
    morekeywords={},
    otherkeywords={\#pragma, omp, parallel,private, shared, schedule, loop, name, master,nowait, reduction, critical, atomic},
    breakatwhitespace=false,         
    breaklines=true,                 
    captionpos=b,                    
    keepspaces=true,                 
    numbers=left,                    
    numbersep=5pt,                  
    showspaces=false,                
    showstringspaces=false,
    showtabs=false,                  
    tabsize=2
}

\usepackage{mfirstuc}
\MFUnocap{are}
\MFUnocap{or}
\MFUnocap{for}
\MFUnocap{a}
\MFUnocap{the}
\MFUnocap{of}
\MFUnocap{to}
\MFUnocap{etc}
\MFUnocap{in}
\MFUnocap{at}
\MFUnocap{and}
\MFUnocap{by}
\MFUnocap{on}
\MFUnocap{that}
\MFUnocap{as}
\MFUnocap{with}
\newcommand{\hide}[1]{}
\newcommand{\REcomment}[1]{}
\renewcommand{\REcomment}[1]{{\textcolor{blue}{\bf [Rudi: #1]}}}

\newcommand{\PBcomment}[1]{}
\renewcommand{\PBcomment}[1]{{\textcolor{orange}{\bf [Parinaz: #1]}}}

\usepackage{array,etoolbox}
\preto\tabular{\setcounter{magicrownumbers}{0}}
\newcounter{magicrownumbers}

\usepackage[hmargin=3cm]{geometry}   
\usepackage{booktabs}
\usepackage{array}
\usepackage{threeparttable}
\usepackage{graphicx}
\newcommand{\tabhead}[1]{\textbf{#1}}

\begin{document}
\title{ \capitalisewords{A comparison between Automatically versus Manually Parallelized NAS Benchmarks
}}
%
\author{Parinaz Barakhshan\inst{1}\orcidID{0000-0001-7232-3923} \and\\
Rudolf Eigenmann\inst{1}\orcidID{0000-0003-1651-827X} }
\authorrunning{P. Barakhshan, R. Eigenmann}
%
\institute{University of Delaware, Newark, DE, USA \\
\email{parinazb@udel.edu} \hspace{0.4cm} 
\url{} 
\email{eigenman@udel.edu}}
\maketitle              
\begin{abstract}
We compare automatically and manually parallelized NAS Benchmarks in order to identify code sections that differ. We discuss opportunities for advancing automatic parallelizers. We find ten patterns that pose challenges for current parallelization technology. We also measure the potential impact of advanced techniques that could perform the needed transformations automatically. While some of our findings are not surprising and difficult to attain -- compilers need to get better at identifying parallelism in outermost loops and in loops containing function calls -- other opportunities are within reach and can make a difference. They include combining loops into parallel regions, avoiding load imbalance, and improving reduction parallelization.   

Advancing compilers through the study of hand-optimized code is a necessary path to move the forefront of compiler research. Very few recent papers have pursued this goal, however. The present work tries to fill this void.


\end{abstract}

\section{Introduction}
\label{Introduction}
Since the end of Dennard scaling~\cite{Dennard} at the turn of the millennium, nearly all computer systems include parallel architectures that are exposed to their programmers. In the past two decades, we have witnessed a significant increase in computer applications in nearly all domains of science, engineering, business, and our daily lives. As a result, the number of program developers has drastically increased, including many software engineers trained on the intricacies of parallel computer architectures and applications, but also an even larger number of non-experts. Tools that help create and efficiently implement parallel applications on modern architectures are more important than ever. While the relevance of
automatic parallelizers is obvious for non-expert programmers, the same tools can also greatly benefit the specialists, assisting them in efficiently performing many of the tedious programming tasks.

After four decades of research in automatic parallelization, a large number of techniques have been developed. Nevertheless, automatic parallelization tools succeed only in about half of today's science and engineering applications. And there is little success in many of the business and daily-life applications, which represent the major part of today's software. Users of parallelizers are often frustrated by the unpredictable performance of automatic tools, which at times degrade the speed below that of the original program. Manual parallelization is often a necessity, but its complexity and tediousness make it amenable to only a minority of highly trained experts. Even for these experts, creating parallel applications is an expensive and time-consuming task. 

Developing tools that automate these tasks is even more challenging. One of the biggest questions is how to bring about advancements in this area. The premise of this
paper is that we need to study representative applications, investigate how manual programmers have performed their tasks, compare the transformations they have applied with those of automatic parallelizers, and learn from these comparisons how to improve our tools. Amazingly, there are very few papers that pursue this
direction. We will discuss these papers in the section on related work.

The present paper tries to fill this void. We identify programming patterns that differ between manually parallelized and auto-parallelized codes, find the limitations of auto-parallelizers,
and suggest improvements for such tools, so that they generate programs that are closer to hand-optimized code.

We do this by studying the NAS Parallel Benchmark (NPB) applications~\cite{npb}. The NPB applications are a representation of real-world applications. While their first release was in 1991, they are continually being modernized and include codes containing irregular code and data patterns. The OpenMP versions of NPB are used as our hand-parallelized applications, which we compare to the serial versions parallelized automatically by the Cetus translator~\cite{CetusCompilerUD}. Cetus is an advanced parallelizer and compiler infrastructure for C programs. We use it to represent modern parallelization technology.

The remainder of the paper is organized as follows. Section~\ref{Scope} outlines our experimental design. Section~\ref{measurements} identifies and measures the code sections that differ between manual and automatic parallelization. Section~\ref{results} presents the main findings, including a description of the code patterns that differ between automatically and manually parallelized applications, an assessment of the performance impact of each pattern, and a discussion of opportunities for compilers. We describe related work in Section~\ref{relatedwork}, followed by conclusions in Section~\ref{conclusion}.

\section{Experimental Design}
\label{Scope}
\paragraph{\textbf{Application Benchmarks:}} 
We use the NAS Parallel Benchmarks NPB 3.3, which includes serial, OpenMP, and MPI codes for ten applications. The original codes are written in Fortran, but we use the variants written in C~\cite{SNUNPB}. We evaluate the codes EP, IS, BT, SP, MG, and CG, which present opportunities for automatic parallelization. For our experiments, we measure the performance of the applications for input Class A, which is a small data set, but representative of larger sets, as we will show in section~\ref{subsec:scale-data}. We report the average performance of three program runs.

\paragraph{\textbf{Automatic Parallelization:}} We use the Cetus open-source automatic parallelizer, which is a source-to-source translator for C programs. Cetus represents some of the most advanced parallelization technology~\cite{Prema2019,Mosseri2020,2022automatic}, including symbolic program analysis. It generates OpenMP-annotated C code on output, invoking GCC as a backend code generator (GCC v4.8.5 with option -O3).
We ran Cetus with its default option to parallelize the codes. Among the major passes applied~\cite{CetusManual} are range analysis, points-to and alias analysis, data dependence analysis, data privatization, induction variable substitution, and reduction parallelization. This experiment uses Cetus as a representative of current auto-parallelizers. Cetus has been actively maintained, with recent improvements~\cite{2022automatic}.

\paragraph{\textbf{Platforms:}}
\label{local}
The key measurements are performed on a four-core system. All CPUs are located on one NUMA node. Each CPU has a 512 KiB L1d cache and a 512 KiB L1i cache, as well as a 4 MiB L2 cache. This system provides full access for easy experimentation; we refer to it as the \textit{Interactive System}.

To validate our findings on a larger system, we also make use of the University of Delaware (UD)'s Caviness Cluster~\cite{Caviness}.  
Caviness is a distributed-memory Linux cluster that was initially deployed in July 2018. A variety of compute nodes are present with different configurations on this cluster. Each node consists of multi-core processors (CPUs), memory, and local disk space. It consists of 126 compute nodes (4536 cores, 24.6 TB memory). The nodes are built of Intel “Broadwell” 18-core processors in a dual-socket configuration for 36 cores per node. Experiments on Caviness need to be submitted via batch queues;
we refer to this cluster as the \textit{Batch System}.

\section{Examining Differences between Manual and Automatic Parallelization}
\label{measurements}
This section presents overall experimental results. We compare the performance of the automatically and manually parallelized applications  (Section~\ref{subsec:overall-interactive}) and identify program sections that exhibit differences (Section~\ref{subsec:program-sections}) between auto- and hand-optimized. We also measure the overheads introduced by program parallelization (Section~\ref{subsec:overhead}).
These measurements were taken on the 4-core Interactive System, introduced in Section~\ref{local}, using data class A. To validate our findings on larger data sets and systems, Section~\ref{subsec:scale-data} and Section~\ref{subsec:scale-cores} also measure and discuss the benchmarks for data class B and up to 36 cores, using the Batch  System described in Section~\ref{local}.

\subsection{\capitalisewords{Performance of Auto-Parallelized and Hand-Parallelized Applications on the interactive system using 1 and 4 cores}}
\label{subsec:overall-interactive}
Table~\ref{table:executionTime1} shows execution times and speedups of the auto-parallelized and hand-parallelized codes, running on the 4-core Interactive System, using the Class A data set.

\vspace*{-\baselineskip}
\begin{table}[H]
\caption{Execution times of the auto- and hand-parallelized codes in seconds. Parallel Execution is measured on 1 and 4 cores on the Interactive System. The parallel speedup is calculated as the run time of the parallel code on 1-core divided by the run time on 4-cores. 
} 
\centering 
\begin{tabular}{|c|c|c|c|c|c|c| } 
\hline\hline 
\toprule
\multicolumn{1}{|c|}{Application} &
\multicolumn{3}{c|}{\textbf{ Auto-Parallelized Code  } } &
\multicolumn{3}{c|}{\textbf{ Manually-Parallelized Code }} \\
{Name}& {Execution Time(s)} & {Execution Time(s)} & {Speedup} & {Execution Time(s)} & {Execution Time(s)} & {Speedup} \\
 & (1 core) &(4 cores)& &(1 core)& (4 cores)& \\
\midrule
\hline 
 SP & 417 &362 &1.2 & 425 &110 &3.8 \\
 BT & 414  &356 &1.2 & 450 &116 &3.8\\
 EP &  86&63  &1.4  & 87 &22  &3.9 \\ 
 MG & 35 &15  &2.3  & 31 &8   &3.8\\ 
 IS & 8  &7   &1.1  & 9 &3   &3.0\\
 CG & 12  &5   &2.4  & 11 &3   &3.7\\ [1ex] 
\hline 
\end{tabular}
\label{table:executionTime1} 
\end{table}

\vspace*{-\baselineskip}

With auto-parallelization, the applications SP, BT, EP, MG, and CG have gained noticeable speedup. For the IS application, there is little gain; the code has not been parallelized substantially due to irregular data patterns. 

The hand-parallelized code yields speedups for all applications. On average, the hand-parallelized codes perform 2.5 times faster than auto-parallelized.
\subsection{\capitalisewords{Performance of Individual Code Sections}}
\label{subsec:program-sections}
Table~\ref{table:ET} on page~\pageref{table:ET} shows the differences between automatically and manually parallelized code sections. 
In addition to the execution times, the table identifies the differing programming patterns.
\begin{table}[H]
  \caption{Differences in individual code sections between auto- and hand-parallelized applications}
  \label{tab:ts}
  \centering
  \begin{threeparttable}
    \begin{tabular}{|p{0.8cm} p{3.1cm} | p{1.5cm} p{1.5cm}| p{0.7cm} | p{0.7cm} | p{0.7cm} | p{0.7cm} | p{0.7cm} |p{0.7cm} | p{0.7cm} | p{0.7cm} |p{0.7cm} | }
      \toprule
       \tabhead{App} &\tabhead{Loop Name \quad \quad } & \tabhead{Auto\tnote{} \quad} & \tabhead{Manual\tnote{} \quad}& \tabhead{P1\tnote{}} & \tabhead{P2\tnote{}} & \tabhead{P3\tnote{}} & \tabhead{P4\tnote{}} &
      \tabhead{P5\tnote{}} &\tabhead{P6\tnote{}} &\tabhead{P7\tnote{}} & \tabhead{P8\tnote{}} & \tabhead{P9\tnote{}} \\
      \midrule
CG  &  main\#1-\#3  & 2.13 & 0.57 & 1 & 0 & 3 & 0 & 0 & 1 & 0 & 0 & 0  \\
 CG  &  conj\_grad\#0-\#4  & 0.15 & 0.15 & 0 & 0 & 5 & 0 & 0 & 0 & 0 & 1 & 1  \\
 CG  &  sparse\#6  & 0.03 & 0.01 & 1 & 0 & 1 & 0 & 0 & 0 & 0 & 0 & 0  \\
 CG  & \textbf{Program}    & 5.00 & 3.00 & 3 & 0 & 8 & 0 & 0 & 1 & 0 & 1 & 1  \\\hline
 IS  &  create\_seq\#0  & 3.48 & 0.88 & 0 & 1 & 1 & 0 & 0 & 1 & 0 & 0 & 1  \\
 IS  &  full\_verify\#0  & 0.12 & 0.05 & 1 & 0 & 1 & 1 & 1 & 1 & 0 & 0 & 1  \\
 IS  &  rank\#1-\#7  & 0.38 & 0.09 & 1 & 0 & 3 & 1 & 1 & 1 & 0 & 0 & 1  \\
 IS  & \textbf{Program}    & 7.39 & 2.80 & 2 & 1 & 5 & 2 & 2 & 3 & 0 & 0 & 3  \\\hline
 MG  &  rprj3\#0  & 0.32 & 0.08 & 1 & 0 & 1 & 0 & 0 & 0 & 0 & 0 & 0 \\
 MG  &  norm2u3\#0   & 0.42 & 0.12 & 1 & 1 & 1 & 0 & 0 & 0 & 0 & 1 & 1 \\
 MG  &  comm3\#0  & 0.003 & 0.003 & 0 & 0 & 3 & 0 & 0 & 0 & 0 & 1 & 1 \\
 MG  &  zran3\#0  & 1.77 & 0.43 & 1 & 1 & 1 & 0 & 0 & 0 & 0 & 0 & 1 \\
 MG  &  zran3\#1-\#3  & 0.25 & 0.03 & 1 & 1 & 3 & 0 & 0 & 0 & 0 & 0 & 1 \\
 MG  & \textbf{Program}  & 15.4 & 8.54 & 4 & 3 & 9 & 0 & 0 & 0 & 0 & 2 & 4 \\\hline
 EP  &  main\#0  & 0.002 & 0.007 & 0 & 0 & 1 & 0 & 0 & 1 & 0 & 0 & 0 \\
 EP  &  main\#3  & 62.5 & 22.4 & 1 & 1 & 1 & 0 & 0 & 1 & 1 & 0 & 1 \\
 EP  & \textbf{Program}    & 63.0 & 22.0 & 1 & 1 & 2 & 0 & 0 & 2 & 1 & 0 & 1 \\\hline
 BT  &  initialize\#0-\#7  & 0.44 & 0.12 & 8 & 7 & 8 & 0 & 0 & 0 & 0 & 6 & 0  \\
 BT  &  exact\_rhs\#0-\#4  & 0.52 & 0.14 & 5 & 3 & 5 & 0 & 0 & 1 & 0 & 2 & 0  \\
 BT  &  compute\_rhs\#0-\#10  & 20.5 & 20.4 & 0 & 0 & 11 & 0 & 0 & 0 & 0 & 7 & 0  \\
 BT  &  x\_solve\#0  & 110 & 31.3 & 1 & 1 & 1 & 0 & 0 & 1 & 0 & 0 & 0  \\
 BT  &  y\_solve\#0  & 110 & 31.5 & 1 & 1 & 1 & 0 & 0 & 1 & 0 & 0 & 0  \\
 BT  &  z\_solve\#0  & 113 & 32.2 & 1 & 1 & 1 & 0 & 0 & 1 & 0 & 0 & 0  \\
 BT  &  error\_norm\#1  & 0.08 & 0.02 & 1 & 1 & 1 & 0 & 0 & 0 & 1 & 1 & 1  \\
 BT  &  rhs\_norm\#1   & 0.004 & 0.003 & 0 & 0 & 1 & 0 & 0 & 0 & 1 & 1 & 1  \\
 BT  & \textbf{Program}    & 356 & 116 & 17 & 14 & 29 & 0 & 0 & 4 & 2 & 17 & 2  \\\hline
 SP  &  error\_norm\#1  & 0.04 & 0.01 & 1 & 1 & 1 & 0 & 0 & 0 & 1 & 1 & 1  \\
 SP  &  rhs\_norm\#1   & 0.003 & 0.002 & 0 & 0 & 1 & 0 & 0 & 0 & 1 & 1 & 1  \\
 SP  &  exact\_rhs\#0-\#4   & 0.77 & 0.13 & 1 & 3 & 5 & 0 & 0 & 1 & 0 & 2 & 0  \\
 SP  &  initialize\#0-\#7  & 0.14 & 0.04 & 1 & 7 & 8 & 0 & 0 & 0 & 0 & 6 & 0  \\
 SP  &  lhsinit\#0  & 0.71 & 0.13 & 1 & 0 & 1 & 0 & 0 & 1 & 0 & 0 & 0  \\
 SP  &  lhsinitj\#0   & 1.10 & 0.30 & 0 & 0 & 0 & 0 & 0 & 1 & 0 & 0 & 0  \\
 SP  &  compute\_rhs\#0-\#10  & 23.3 & 20.1 & 0 & 0 & 11 & 0 & 0 & 0 & 0 & 7 & 0  \\
 SP  &  x\_solve\#0  & 87.2 & 20.4 & 1 & 1 & 1 & 0 & 0 & 1 & 0 & 0 & 0  \\
 SP  &  y\_solve\#0  & 123 & 20.8 & 1 & 1 & 1 & 0 & 0 & 1 & 0 & 0 & 0  \\
 SP  &  z\_solve\#0  & 123 & 21.4 & 1 & 1 & 1 & 0 & 0 & 1 & 0 & 0 & 0 \\
 SP  & \textbf{Program}  & 362 & 111 & 7 & 14 & 30 & 0 & 0 & 6 & 2 & 17 & 2 \\
      \bottomrule
    \end{tabular}
    \begin{tablenotes}
      \item Auto-- Execution time of auto-parallelized code (seconds) on 4 cores 
      \item Manual-- Execution time of manually-parallelized code (seconds) on 4 cores
      \item P1-- Number of loops in which the outermost loop is parallelized; more on section~\ref{Parallelizingnestedloops} 
      \item P2-- Number of loops containing function calls inside the region; more on section~\ref{functioncalls}     
      \item P3-- Number of loops inside the parallel region; more on section~\ref{ParallelRegionPattern}     
      \item P4-- Dynamic schedule (1 means the pattern is applied); more on section \ref{DynamicScheduling}
      \item P5-- Irregular patterns like indirect array accesses (1 means the pattern is applied); more on section~\ref{IrregularPatterns}
      \item P6--  Threadprivate data access (1 means Threadprivate data has been accessed); more on section~\ref{threadprivate}
      \item P7-- Array reduction (1 means the pattern is applied); more on section~\ref{ArrayReductionPattern}
      \item P8-- Number of NOWAIT clauses; more on section~\ref{NOWAITPattern}
      \item P9-- Code modification (1 means the pattern is applied); more on section~\ref{CodeModifications}

    \end{tablenotes}
  \end{threeparttable}
  \label{table:ET}
\end{table}
We number adjacent (nests of) loops in each subroutine from 0. For example, main\#1-\#3 indicates the second through the fourth \textit{for loops} in function {\em main}.

Our comparison of auto- and hand-parallelized codes revealed several differing program patterns. The table shows which of these patterns have been used in the manually parallelized codes. Section~\ref{results} explains each of these patterns in more detail and quantifies  the performance differences they make. We omit code sections with insignificant performance differences.

We examine the reasons behind the differences between the auto-parallelized and hand-parallelized execution times in Section~\ref{results} and discuss the programming patterns responsible for these differences.

\subsection{\capitalisewords{Overhead of parallel transformations in hand-parallelized applications}}
\label{subsec:overhead}
The inserted OpenMP directives (also called compiler pragmas) determine how the compiler and run-time system transform the code, manage and synchronize teams of threads and distribute work among the threads. Table~\ref{table:OMP} shows the serial and 1-core parallel execution of the hand-parallelized applications, exhibiting the introduced overhead. Two of the six applications incur overheads of more than 3\% while in two of the applications the performance improves by 9\% and 12\%. Some of the sources of overhead are:  

\begin{table}[H]
\caption{Parallel overhead in manually optimized codes} 
\centering 
\begin{tabular}{| c c c c|} 
\hline\hline 

\multicolumn{1}{|p{2cm}|}{\centering Application Name}
& \multicolumn{1}{|p{3cm}|}{\centering Serial Execution Time (s)  }
& \multicolumn{1}{|p{4cm}|}{\centering Execution Time of hand-parallelized code(s) on 1 core }
& \multicolumn{1}{|p{2cm}|}{\centering Difference}
\\\hline
\hline 
 SP & 416 \hide{& 417}  &  425 & 3\%\\
 BT & 414 \hide{& 414}  &  450 & 9\%\\
 EP & 86 \hide{& 86}   &  87  & 1\%\\ 
 MG & 35 \hide{& 35}   &  31  & -12\%\\ 
 IS & 8 \hide{& 8}    &  9   & 11\%\\
 CG & 12 \hide{& 12}   &  11  & -9\%\\ [1ex] 
\hline 
\end{tabular}
\label{table:OMP} 
\end{table}

\begin{itemize}
    \item  OpenMP in such compilers as GCC, LLVM, and Intel ICC is implemented using outlining~\cite{outliningOpenmp}, extracting the parallel region into its own function. 
    The extra function call, initiating communications with the helper threads, copying shared variables to the helpers, assigning work shares, as well as synchronizing the threads, all contribute to this overhead. 
    \item In addition, we found that parallel regions that use thread-private data may incur extra overheads, as described in more detail in Section~\ref{threadprivate}. 
    \item Moreover, the addition of OpenMP directives can sometimes prevent the compiler from performing certain optimizations. 
\end{itemize}

\subsection{\capitalisewords{Application scalability with increasing data set and system performance}}
\label{subsec:scale-data}
To verify the validity of our findings on larger systems and data sets, we executed our benchmarks on the Batch System described in Section~\ref{local}. We used the same core counts as in the prior sections (1 and 4 cores) but included in our measurements data set Class B, next to Class A.

Tables~\ref{table:executionTime2} and~\ref{table:executionTime3} show the code performance on the Batch System using Class A and Class B data sets, respectively. Using Class A, the hand-parallelized codes show an average 4-core speedup of 3.6. The corresponding gain for the auto-parallelized codes is 1.9.

Using the larger class-B data set, the average speedups are close to those for Class A -- 3.7 for manually parallelized and 2.1 for automatically parallelized. While the hand-parallelized codes achieve a speedup close to the available core count, four out of the six auto-parallelized codes yield substantially lower performance. These findings match the conclusions obtained for Class A on the Interactive System, justifying the use of the more nimble Interactive System and Class A for our detailed experiments.
\vspace*{-\baselineskip}
\begin{table}[H]
\caption{Performance of auto- and hand-parallelized codes in seconds. Parallel Execution is measured on 1 and 4 cores. Parallel Speedup is calculated as the parallel run-time on 1-core divided by the run-time on 4-cores. The measurements use \textbf{Class A data set and the Batch System}.} 
\centering 
\begin{tabular}{|c|c|c|c|c|c|c| } 
\hline\hline 
\toprule
\multicolumn{1}{|c|}{Application} &
\multicolumn{3}{c|}{\textbf{  Auto-Parallelized Code  } } &
\multicolumn{3}{c|}{\textbf{ Manually-Parallelized Code }} \\
{Name}& {Execution Time(s)} & {Execution Time(s)} & {Speedup} & {Execution Time(s)}& {Execution Time(s)} & {Speedup}\\
& (1 core) &(4 cores)& &(1 core)& (4 cores)&\\
\midrule
\hline 
 SP &134 &72 &1.9 &133 &37 &3.6 \\
 BT &146 &126 &1.2 &155 &43 &3.6\\
 EP &33 &23  &1.4  &33 &8.8  &3.7\\
 MG &5.7 &2.1 &2.7 &5.7 &1.5   &3.7\\
 IS &0.8  &0.8  &1.0  &0.7 &0.2 &3.1\\
 CG &2  &0.5  &3.7  &2 &0.5  &3.7\\ [1ex] 
\hline 
\end{tabular}
\label{table:executionTime2} 
\end{table}

\vspace*{-\baselineskip}
\begin{table}[H]
\caption{Performance of auto- and hand-parallelized codes in seconds. Parallel Execution is measured on 1 and 4 cores. Parallel Speedup is calculated as the parallel run-time on 1-core divided by the run-time on 4-cores. The measurements use \textbf{Class B data set and the Batch System}.} 
\centering 
\begin{tabular}{|c|c|c|c|c|c|c| } 
\hline\hline 
 \toprule
\multirow{1}{*}{Application} &
      \multicolumn{3}{c|}{\textbf{  Auto-Parallelized Code  } } &
      \multicolumn{3}{c|}{\textbf{ Manually-Parallelized Code }} \\
{Name}&{Execution Time(s)} & {Execution Time(s)} & {Speedup} & {Execution Time(s)}& {Execution Time(s)} & {Speedup} \\
 & (1 core) &(4 cores)& &(1 core)& (4 cores)&\\
\midrule
\hline 
 SP &543 &284 &1.9 &556 &151 &3.7 \\
 BT &595 &518 &1.1 &638 &169 &3.8\\
 EP &133 &94  &1.4 &129 &35 &3.7\\
 MG &26  &8.5 &3.1 &25.5 &7 &3.7\\
 IS &3.4 &3   &1.0 &3.2  &0.9 &3.6\\
 CG &84  &22  &3.9 &80 &22  &3.7\\ [1ex] 
\hline 
\end{tabular}
\label{table:executionTime3} 
\end{table}

\subsection{\capitalisewords{Application Scalability with increasing core counts using Class B Data set}}
\label{subsec:scale-cores}
The previous section shows good efficiency of the hand-parallelized codes on four cores. The auto-parallelized versions execute at about half that performance, on average. This section tests the scalability of programs to up to 36 cores, using the Class B data set.

\vspace*{-\baselineskip}
\begin{figure}[h]
\centering
\begin{subfigure}{.5\textwidth}
  \centering
  \hspace*{-0.4cm}\includegraphics[width=1\linewidth, height=5.5cm]{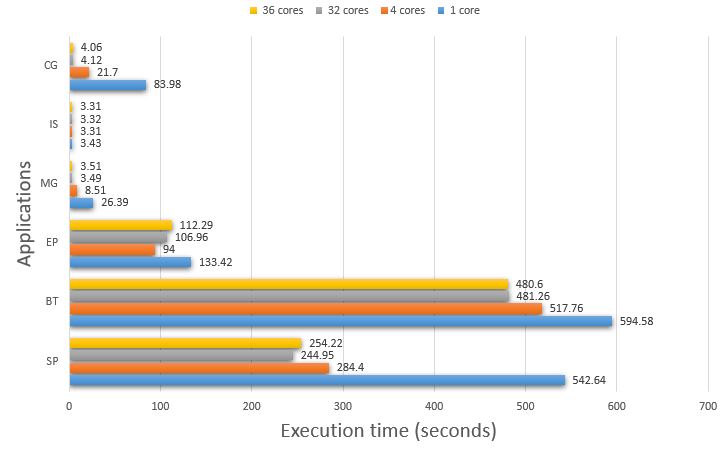}
  \caption{Auto-parallelized codes }
  \label{fig:ETAuto}
\end{subfigure}%
\begin{subfigure}{.5\textwidth}
  \centering
  \hspace*{-0cm}\includegraphics[width=1\linewidth, height=5.5cm]{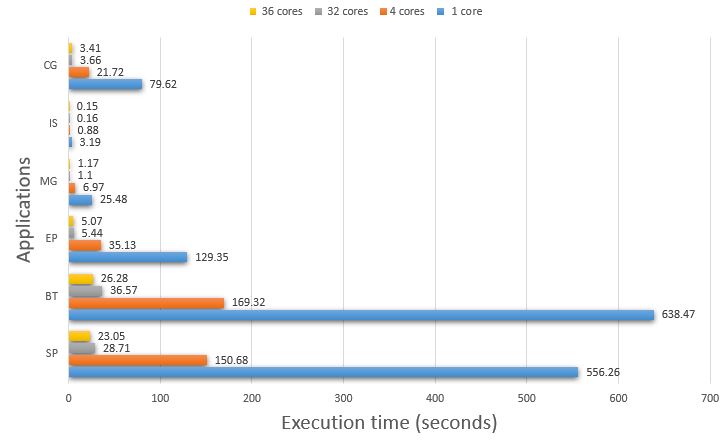}
  \caption{Hand-parallelized codes }
  \label{fig:ETHand}
\end{subfigure}
\caption{Comparing the execution time of auto-parallelized and hand-parallelized codes of Class B data-set using different core counts }

\label{fig:ET}
\end{figure}

\vspace*{-\baselineskip}
\begin{figure}[H]
\centering
\begin{subfigure}{.5\textwidth}
  \centering
  \includegraphics[width=1\linewidth, height=5cm]{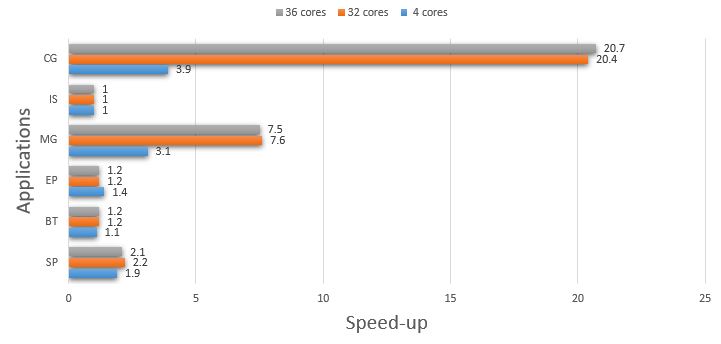}
  \caption{Auto-parallelized codes}
  \label{fig:SpeedUpAuto}
\end{subfigure}%
\begin{subfigure}{.5\textwidth}
  \centering
  \includegraphics[width=1\linewidth, height=5cm]{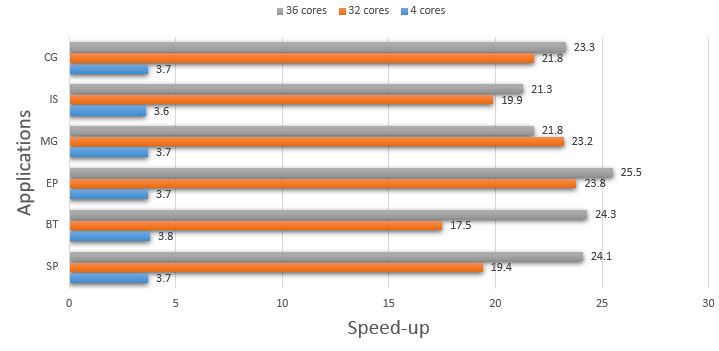}
  \caption{Hand-parallelized codes}
  \label{fig:SpeedUpHand}
\end{subfigure}
\caption{Comparing the speedup of the auto-parallelized and hand-parallelized codes of Class B data-set on different core counts }
\label{fig:SpeedUP}
\end{figure}
Figures \ref{fig:ET} and \ref{fig:SpeedUP} 
show the code execution times and speedups, respectively, with increasing core counts. Auto-parallelized codes cannot efficiently utilize the available cores. Except for CG, and MG, adding more than 4 cores does not increase the speed. When increasing the number of cores beyond 32, the performance gain is not significant and may even worsen. For the hand-parallelized codes, as the number of cores increases from 32 to 36, the performance continues to improve, except for MG. In general, the hand-parallelized benchmarks make efficient use of the available cores. 

Overall, the qualitative findings of the 4-core measurements remain the same: While the hand-parallelized codes show good scalability, the performance of the auto-parallelized versions is limited. That limitation increases with higher core counts. Hence the importance of the techniques presented in the next section, which aim to bring the efficiency of the auto-parallelized codes closer to manually parallelized, will be even larger than shown by the 4-core evaluation.

\section{Code Patterns, Performance Impact, Opportunities for Compilers}
\label{results}
We now analyze the program sections in each benchmark that differ between the manually and automatically parallelized versions. We have identified ten programming patterns that represent these differences. In rough order of importance, we first explain the pattern, followed by assessing the performance impact of enabling/disabling the pattern. We then discuss the potential for improving compilers to implement the pattern. We quantify the potential impact of these techniques on our 4-core Interactive System. As mentioned at the end of Section~\ref{measurements}, these numbers represent lower bounds, which will increase for systems with higher core counts.

\subsection{ \capitalisewords{Parallelizing nested loops at the outermost level possible}}
\label{Parallelizingnestedloops}
\subsubsection{The Pattern} It is well understood that outermost parallelism yields the best performance and that automatic parallelization may fail to do so, finding parallelism in inner loops, only. Running outermost loops in parallel minimizes the number of parallel loop invocations, and the associated fork/join overheads, including implicit barriers at the loop end. Not surprisingly, we found several program sections that differed between manually and automatically parallelized in this way. Among the causes were irregular data accesses, function calls (discussed later), or dependences that could not be disproven.

\subsubsection{\textbf{Performance Impact} }
Subroutines x\_solve(), y\_solve(), and z\_solve() in programs BT and SP are examples of compute-intensive functions that are not parallelized at the outermost level by the Cetus compiler --  due to function calls present in the outermost loop. Table~\ref{table:OuterMostLoop} shows the differences between the auto- and hand-parallelized code execution times. 
\begin{table}[H]
\caption{Impact of parallelizing the outermost loop in nested loops, comparing the execution time of the Cetus-parallelized code with the manually parallelized code.} 
\centering 
\begin{tabular}{ | c c c c c |} 
\hline\hline 
 \multicolumn{1}{|p{3cm}|}{\centering Application Name}
& \multicolumn{1}{|p{2cm}|}{\centering Loop Name }
& \multicolumn{1}{|p{3.5cm}|}{\centering Technique Not Applied }
& \multicolumn{1}{|p{3cm}|}{\centering Technique Applied }
& \multicolumn{1}{|p{2cm}|}{\centering Impact}\\\hline
\hline 
 BT & x\_solve\#0  & 110 & 31 &   255\%\\
 BT  & y\_solve\#0 & 110 & 31 &   255\%\\
 BT  & z\_solve\#0 & 113  & 32   &   253\%\\
 BT  & \textbf{program} & 356  & 116  &  206\%\\
 SP  & x\_solve\#0     & 87 &  20  &   335\%\\
 SP  & y\_solve\#0    & 123 &  21  &   486\%\\
 SP  & z\_solve\#0    & 123 & 21   &   486\%\\
 SP  & \textbf{program }   & 362 &  111  & 226\%\\
[1ex] 
\hline 
\end{tabular}
\label{table:OuterMostLoop} 
\end{table}

\subsubsection{\textbf{Opportunities for Compilers} }
The presence of function calls requires inter-procedural analysis or inline expansion capabilities, which we will discuss in the following subsection. Irregular access patterns have long been a challenge for compilers, with both run-time and recent compile-time approaches pursuing improvements. For disproving data dependences, we have often found that the opportunity is in the propagation of information across the program (such as interprocedural symbolic analysis) rather than in increasing the power of data dependence tests themselves.

\subsection{\capitalisewords{Parallelizing loops containing function calls}}
\label{functioncalls}
\subsubsection{\textbf{The Pattern} }Most auto-parallelizers, including Cetus, do not consider loops with function calls or I/O statements for parallelization, unless those functions are known to be side-effect free. Our study found many examples in which a function call inside a loop prevented auto-parallelization. The same loops were parallelized in the manually parallelized codes. 

Inline expansion, which replaces a function call with the body of the called subroutine, can help parallelize such patterns. Users of the Cetus compiler have that option available. We will measure the effect of doing so, next.

\subsubsection{\textbf{Performance Impact} }
We performed an experiment to determine how much parallelization is enabled through inline expansion in Cetus-parallelized codes. Table~\ref{table:Inlining}  
shows the result.

\begin{table}[H]
\begin{threeparttable}[b]
\caption{Impact of inlining in parallelizing automatically parallelized codes; Comparing the execution time of auto-parallelized codes before and after inlining.} 
\centering 
\begin{tabular}{ |c c H c c c c H c|} 
\midrule
\hline\hline 

 \multicolumn{1}{|p{1cm}|}{\centering App Name}
& \multicolumn{1}{|p{2cm}|}{\centering Loop Name }
& \multicolumn{1}{H}{\centering Number of Loops Containing Function Call }
& \multicolumn{1}{|p{2.5cm}|}{\centering Number of Inlined Functions }
& \multicolumn{1}{|p{2cm}|}{\centering Parallelized after Inlining? }
& \multicolumn{1}{|p{2.5cm}|}{\centering Technique Not Applied }
& \multicolumn{1}{|p{2.5cm}|}{\centering Inlining Technique Applied}
& \multicolumn{1}{H}{\centering Performance Difference }
& \multicolumn{1}{|p{1cm}|}{\centering Impact}\\\hline
\midrule
\hline 
 BT & initialize\#0-\#7   & 7 &9 & Yes$^1$ & 0.44   & 0.14  & &214\%\\
 BT &  exact\_rhs\#0-\#4  & 3 & 3 & Yes$^2$ & 0.52  & 0.63  & &-17\%\\
 BT & x\_solve\#0 & 1         & 8 & No  & 110 & 122  & &-10\%\\
 BT  & y\_solve\#0    & 1     & 8 & No  & 110 &  123 & &-11\%\\
 BT  &  z\_solve\#0 &1       & 8 & No  & 113 & 124  & &-9\%\\
 BT  &  error\_norm\#1    & 1 & 1& Yes$^3$ &  0.08  &  0.03 & &167\%\\
 BT  & \textbf{Program}   &   & &  & 356 & 395  & 0.903 & -10\%\\
 SP & initialize\#0-\#7   & 7 & 9& Yes$^1$ & 0.14  & 0.04  & &250\%\\
  SP &  exact\_rhs\#0-\#4  & 3  & 3 & Yes$^2$ & 0.77 & 0.7  & & 10\%\\
 SP & x\_solve\#0 & 1        &1 & No  & 87 & 87  & & 0\%\\
 SP &  y\_solve\#0      & 1  &1 & No  & 123 & 124  & & -1\%\\
 SP &  z\_solve\#0 &1       & 1 & No  & 123 &  123 & & 0\%\\
  SP &  error\_norm\#1    & 1 &1 & Yes$^3$ &  0.04 & 0.03 & & 33\%\\
 SP & \textbf{Program}    & &  & & 362 & 377  &  & -4\%\\
[1ex] 
\hline 
 \midrule 
\end{tabular}
\label{table:Inlining} 
    \begin{tablenotes}
       \item [1] In nested loop structures, the outermost loops are parallelized. In manually-parallelized code, however, all parallel loops are included in a parallel region.
       \item [2] In nested loop structures, inner loops are parallelized.
       \item [3] While the outermost loop is parallelized, the array reduction implementation differs from the hand-parallelized code that will be discussed later.
     \end{tablenotes}
\end{threeparttable}
\end{table}
We found that auto-parallelization indeed could detect additional parallelism in several of the loops in question after applying inlining. As displayed in Table~\ref{table:Inlining}, subroutine {\em initialize()} in both BT and SP shows significant performance gain due to parallelization of the outermost loops. However, in {\em exact\_rhs()}, the transformation led to performance degradation. While additional loops could be parallelized, these were inner loops where parallelization was not profitable. What's more, the most compute-intensive loops, in subroutines x\_solve(), y\_solve(), and z\_solve() of both applications, remained unaffected by inline expansion, as Cetus is still unable to disprove data dependences. 

\subsubsection{\textbf{Opportunities for Compilers} }
Despite studies on interprocedural analysis (IPA) that have been carried out for more than three decades, IPA is not available in most compilers. Among the reasons are the complexity of the technique, the fact that most analyses need specialized IPA algorithms, and the resulting increase in compilation times.

By interacting with the user, it is possible to identify user functions that have no side effects and add them to the default list of side-effect-free functions, which consist primarily of math functions. 
Other opportunities include selective subroutine inline expansion during the compiler analysis only. The former technique could identify additional parallelism with user input, while the latter would eliminate overheads, such as excessive code growth.

\subsection{\capitalisewords{Parallel regions enclosing multiple Parallel loops}}
\label{ParallelRegionPattern}
\subsubsection{\textbf{The Pattern} }In OpenMP, multiple adjacent parallel loops can be converted into a {\em parallel region}. This way, the parallel threads are spawned only once, at the beginning of the region, reducing fork/join overhead. The original parallel loops will become {\em worksharing constructs}, which simply distribute their iterations onto the available threads. In some cases, the programmers had inserted {\em NOWAIT} clauses to eliminate barrier synchronizations at the end of the worksharing constructs.
In the hand-parallelized codes, we found this pattern frequently. By contrast, auto-parallelizers, including Cetus, typically examine and parallelize loops individually.
 
\subsubsection{\textbf{Performance Impact} }
We have measured the impact of such a technique by converting the hand-parallelized programs to variants without parallel regions. The loops inside the regions were changed to individual parallel loops. Note that doing so also forces a barrier synchronization at the end of each parallel loop. The results are presented in  Table~\ref{table:ParallelRegion}. 

\vspace*{-\baselineskip}
\begin{table}[H]
\caption{Impact of enclosing multiple parallel loops in a parallel region -- comparing the execution times of the code containing individual parallel loops with the hand-optimized code containing a parallel region.} 
\centering 
\begin{tabular}{| c c c c c H c|} 
\hline\hline 

 \multicolumn{1}{|p{2cm}|}{\centering App Name}
& \multicolumn{1}{|p{2.5cm}|}{\centering Loop Name }
& \multicolumn{1}{|p{2.5cm}|}{\centering Number of Loops }
& \multicolumn{1}{|p{3.5cm}|}{\centering Technique Not Applied}
& \multicolumn{1}{|p{2.8cm}|}{\centering Technique Applied }
& \multicolumn{1}{H}{\centering Performance Difference }
& \multicolumn{1}{|p{1cm}|}{\centering Impact }\\\hline
\hline 
 MG  &  comm3\#0-\#2  &3   &  0.003 & 0.003 & 1 & 0\%\\
  MG  &  zran3\#1-\#3  &3   &  0.034 & 0.033 & 0.247 & 4\%\\
 MG &   \textbf{Program} &6   &  9 & 8.5 & 0.969 & 6\%\\

 BT & initialize\#0-\#7   & 8 & 0.124 &  0.116  & 0.938 & 7\%\\
 BT &  exact\_rhs\#0-\#4    & 5  & 0.167 & 0.142  & 0.850 & 18\%\\
 BT &  compute\_rhs\#0-\#10  & 11  &  0.117& 0.108  & 0.926 & 8\%\\
 BT  & \textbf{Program}  & 24  &  129 & 116  &  0.904 & 11\%\\[1ex]

\hline 
\end{tabular}
\label{table:ParallelRegion} 
\end{table}

\subsubsection{\textbf{Opportunities for Compilers} }
Developing transformations that combine adjacent parallel loops into a parallel region seems feasible in some situations, but we are not aware of auto-parallelizers that do so. In other cases, creating parallel regions can be challenging because  sequential code sections may be present between the parallel loops. There exists work on eliminating barrier synchronization, which can be incorporated into such techniques.

\subsection{\capitalisewords{Avoiding load Imbalance through dynamic scheduling}}
\label{DynamicScheduling}
\subsubsection{\textbf{The Pattern} } Load imbalance can be caused by the uneven distribution of work across worker threads. Loop scheduling defines chunks of loop iterations and their distribution onto the threads. In loops that are prone to uneven workload, due to conditional execution or work that depends on the iteration number, loop scheduling can affect performance noticeably.
Two schedule clauses offered by OpenMP for resolving load imbalance are {\em dynamic} and {\em guided}. They make scheduling decisions at run-time, assigning chunks of iterations to idle threads. 

The developers of the hand-parallelized codes have made use of these clauses. By contrast, the Cetus compiler currently does not change the default loop schedule, which is the static distribution of an equal share of iterations to all worker threads.


 

\subsubsection{\textbf{Performance Impact} }
We have found some loops where loop scheduling made a substantial difference. Table~\ref{table:DynamicSchedule}, shows two such loops in the IS program. The improved code performance of 43\% and 17\%, respectively, translates to a  noticeable overall program speed improvement of 6\%. The impact of dynamic scheduling on the whole application is significant, as the rank function is invoked multiple times.

\begin{table}[H]
\caption{Impact of adding dynamic scheduling. The execution time of the code when scheduling is disabled is compared to the execution time of the  manually parallelized code.} 
\centering 
\begin{tabular}{| c c c c H c|} 
\hline\hline 

 \multicolumn{1}{|p{3cm}|}{\centering Application Name}
& \multicolumn{1}{|p{2.5cm}|}{\centering Loop Name }
& \multicolumn{1}{|p{3.5cm}|}{\centering Technique Not Applied}
& \multicolumn{1}{|p{3cm}|}{\centering Technique Applied}
& \multicolumn{1}{H}{\centering Performance Difference }
& \multicolumn{1}{|p{2cm}|}{\centering Impact }\\\hline
\hline 
 IS & full\_verify\#0   & 0.07 & 0.05 & 0.700 & 43\%\\
 IS & rank\#1-\#7       & 0.10 & 0.09 & 0.854  & 17\%\\
 IS & \textbf{program}            & 2.14 & 2.02  & 0.944 & 6\%\\ [1ex] 
\hline 
\end{tabular}
\label{table:DynamicSchedule} 
\end{table}
\subsubsection{\textbf{Opportunities for Compilers} }
Program and loop workloads are affected by both program and execution characteristics. Dynamic factors, such as external programs in shared machines, and conditional executions guided by input data, are difficult to assess. However, the compiler {\em can} analyze programs for conditional execution patterns that may depend on input data, iteration numbers that tend to load threads unevenly, and inner loops whose workload depends on outer loop iterations (e.g., triangular loops).

\subsection{\capitalisewords{analyzing irregular data access patterns in hand-parallelized codes}}
\label{IrregularPatterns}
\subsubsection{\textbf{The Pattern} }Applications that have irregular data access patterns with complex code structures prevent auto-parallelizers from succeeding. 

The IS application exhibits such patterns. The loops full\_verify\#0, rank\#0, rank\#2, rank\#4, and rank\#6 include indirect array accesses, which prevents Cetus from detecting parallelism. 

\subsubsection{\textbf{Performance Impact} }
Table~\ref{table:irregular} reports execution times of these loops in the IS application when they are not parallelized by the auto-parallelizer due to such patterns, and when they are parallelized in the manually parallelized code.

\begin{table}[H]
\caption{Impact of parallelizing irregular patterns. The execution time of the auto-parallelized code, where irregular patterns remain serial, is compared to the execution time of the manually parallelized code, where the same loops are parallelized.} 
\centering 
\begin{tabular}{| c c c c c|} 
\hline\hline 

 \multicolumn{1}{|p{3cm}|}{\centering Application Name}
& \multicolumn{1}{|p{2cm}|}{\centering Loop Name }
& \multicolumn{1}{|p{3.5cm}|}{\centering Technique Not Applied}
& \multicolumn{1}{|p{3cm}|}{\centering Technique Applied}
& \multicolumn{1}{|p{2cm}|}{\centering Impact}\\\hline
\hline 
IS & full\_verify\#0 & 0.12 & 0.05 & 135\%\\
IS & rank\#1-\#7 & 0.38 & 0.09 & 318\%\\
IS  & \textbf{program }   & 7.39 &  2.80  &  163\%\\
[1ex] 
\hline 
\end{tabular}
\label{table:irregular} 
\end{table}

\subsubsection{\textbf{Opportunities for Compilers} }
Loops containing subscripted subscripts are among the most complex patterns for compilers to analyze. A number of run-time techniques have been developed, such as run-time data-dependence tests and inspector-executor schemes. Recent work has also begun to develop compile-time techniques based on the observation that, in some cases, the information needed to prove the absence of data dependences is present in the application program~\cite{2022automatic}~\cite{bhosale2021automatic}.

\subsection{\capitalisewords{Threadprivate Data}}
\label{threadprivate}
\subsubsection{\textbf{Pattern Explanation} }The OpenMP threadaprivate directive specifies that variables are replicated, with a copy being kept in each thread. It privatizes static or global variables that are modified by multiple parallel regions. Threadprivate variables persist across regions. The manually parallelized benchmarks make use of this concept in a number of program sections.

Auto-parallelizers, including Cetus, do not create threadprivate data. Data that need to be replicated across threads and persist across parallel regions or loops need to be implemented through data expansion or copying region/loop-private data in and out -- sometimes through first/last-private clauses. 

\subsubsection{\textbf{Performance Impact} }
We measured the impact of using threadprivate data by considering those program sections where conversion to loop-private data was possible. We compared the performance of the variants without threadprivate data (loop-private data only) with the hand-parallelized variants, which use threadprivate data.

The result was unexpected. Table~\ref{table:Threadprivatization} on page~\pageref{table:Threadprivatization} shows that using threadprivate data lowers the performance in all of our cases. The compute-intensive loops in BT, subroutine x/y/z\_solve, see a 25\% performance reduction. The superior performance of regions without the use of threadprivate data is consistent with the findings of others~\cite{threadprivate}, who attribute this effect to inefficient  OpenMP implementations.

We did not measure other program sections where additional programming steps would be necessary for the transformation to region/loop-private data. In these cases, the additional steps would likely add overhead, making the threadprivate variant more desirable.

\begin{table}[H]
\caption{Impact of using Threadprivate directives. We compare the execution time of the code where threadprivate data is replaced by loop-private data, with the execution time of the  manually parallelized code.} 
\centering 
\begin{tabular}{| c c c c H c|} 
\hline\hline 
 \multicolumn{1}{|p{3cm}|}{\centering Application Name}
& \multicolumn{1}{|p{2.5cm}|}{\centering\arraybackslash Loop Name }
& \multicolumn{1}{|p{3.5cm}|}{\centering Technique Not Applied}
& \multicolumn{1}{|p{3cm}|}{\centering Technique Applied}
& \multicolumn{1}{H}{\centering Performance Difference }
& \multicolumn{1}{|p{2cm}|}{\centering Impact }\\\hline
\hline 
 EP & main\#0  & 0.003 & 0.006  & 2 & -50\%\\
 EP & main\#3  & 20.93 & 22.40 & 1.072  & -7\%\\
 EP & \textbf{Program }  & 21.0 & 22.5 & 1.072  & -7\%\\
 BT & exact\_rhs\#0-\#4   & 0.055 & 0.142 & 2.083  & -61\%\\
 BT & x\_solve\#0   & 23.43 & 31.24 & 1.333 & -25\%\\
 BT & y\_solve\#0   & 23.63 & 31.51 & 1.333 & -25\%\\
 BT & z\_solve\#0   & 24.45 & 32.17 & 1.315 & -24\%\\
 BT & \textbf{ Program  }   & 93 & 116 & 1.256  & -20\%\\[1ex] 
\hline 
\end{tabular}
\label{table:Threadprivatization} 
\end{table}

\subsubsection{\textbf{Opportunities for Compilers} }
Identifying threadprivate variables would involve analyses similar to current data privatization, combined with liveness analysis across loops. While this appears feasible, careful consideration will need to be given to profitability, so as to avoid situations with negative impacts.

\subsection{\capitalisewords{Array Reductions }}
\label{ArrayReductionPattern}
\subsubsection{\textbf{The Pattern} }
Reduction operations are parallelized as follows: Each thread concurrently performs a reduction operation on the assigned loop iterations, creating partial results, followed by a step that combines the partial results. We have found differences in this combination step. For array reductions, the hand-parallelized versions perform the needed mutual exclusion operation on each element individually, using an OpenMP {\em atomic} construct. By contrast, the Cetus implementation performs the mutual exclusion for the entire array, by means of a {\em critical section}. This is the major difference, next to a minor variation in how data is allocated. 
 
The following example is taken from the BT, rhs\_norm() subroutine. A reduction optimization can be applied to the {\tt rms} array shown in Listing~\ref{ex14}, on line 6, using a (+) reduction operator.

\begin{lstlisting}[language=C , caption=Example taken from the BT\, rhs\_norm() procedure in which an array reduction technique can be applied, label=ex14]
  for (k = 1; k <= grid_points[2]-2; k++) {
    for (j = 1; j <= grid_points[1]-2; j++) {
      for (i = 1; i <= grid_points[0]-2; i++) {
        for (m = 0; m < 5; m++) {
          add = rhs[k][j][i][m];
          rms[m] = rms[m] + add*add;
        } 
      } 
    } 
  }
\end{lstlisting}

Listing~\ref{ex15} on page~\pageref{ex15} shows how Cetus implements array reduction. 
On line 3, the temporary reduction array, \emph{reduce}, is dynamically allocated in shared space and initialized by the loop on line 5. The array is used on line 20 to perform the partial reduction.
On line 26, a {\em Critical} construct is used to protect the concurrent update of \emph{rms} by multiple threads. The critical section encloses a loop that will be executed by each thread sequentially.

\begin{lstlisting}[language=C , caption=Cetus implementation of the array reduction, label=ex15]
	#pragma omp parallel if((10000<(((((((-43L+(92L*grid_points[0L]))+(86L*grid_points[1L]))+(89L*grid_points[2L]))+((-46L*grid_points[0L])*grid_points[1L]))+((-46L*grid_points[0L])*grid_points[2L]))+((-43L*grid_points[1L])*grid_points[2L]))+(((23L*grid_points[0L])*grid_points[1L])*grid_points[2L])))) private(add, i, j, k, m)
	{
		double * reduce = (double * )malloc(5*sizeof (double));
		int reduce_span_0;
		for (reduce_span_0=0; reduce_span_0<5; reduce_span_0 ++ ) {
			reduce[reduce_span_0]=0;
		}
		
		#pragma loop name rhs_norm#1 
		#pragma omp for
		for (k=1; k<=(grid_points[2]-2); k ++ )	{
			#pragma loop name rhs_norm#1#0 
			for (j=1; j<=(grid_points[1]-2); j ++ )	{
				#pragma loop name rhs_norm#1#0#0 
				for (i=1; i<=(grid_points[0]-2); i ++ )	{
					#pragma loop name rhs_norm#1#0#0#0 
					for (m=0; m<5; m ++ )
					{
						add=rhs[k][j][i][m];
						reduce[m]=(reduce[m]+(add*add));
					}
				}
			}
		}
		
		#pragma omp critical
		{
			for (reduce_span_0=0; reduce_span_0<5; reduce_span_0 ++ ) {
				rms[reduce_span_0]+=reduce[reduce_span_0];
			}
		}
	}//end of parallel region
\end{lstlisting}

Listing\ref{ex16} on page~\pageref{ex16} demonstrates how array reduction is implemented in the hand-parallelized code in the same example. This implementation differs from the previous implementation in the following ways:
\begin{itemize}
    \item Using a privatized version of the reduction array {\em rms\_local}, instead of a dynamically allocated array.
    \item On line 7, a NOWAIT clause is used to omit the barrier at the end of the for loop, allowing each thread to immediately proceed to the update step.
    \item Using an {\em atomic} construct inside the for loop on line 18 to combine the partial results. 
\end{itemize}
\vspace*{-\baselineskip}
\begin{lstlisting}[language=C , caption=Array reduction implemented in hand-parallelized code, label=ex16]
  double rms_local[5];
  #pragma omp parallel default(shared) private(i,j,k,m,add,rms_local) shared(rms)
  {
  for (m = 0; m < 5; m++) {
    rms_local[m] = 0.0;
  }
  #pragma omp for nowait
  for (k = 1; k <= grid_points[2]-2; k++) {
    for (j = 1; j <= grid_points[1]-2; j++) {
      for (i = 1; i <= grid_points[0]-2; i++) {
        for (m = 0; m < 5; m++) {
          add = rhs[k][j][i][m];
          rms_local[m] = rms_local[m] + add*add;
        } 
      } 
    } 
  } 
  for (m = 0; m < 5; m++) {
    #pragma omp atomic
    rms[m] += rms_local[m];
  }
  } //end of parallel region
\end{lstlisting}

\subsubsection{\textbf{Performance Impact} }
We compared the two variants, individual element synchronization (manual) and overall synchronization (automatic), by replacing two of the array reduction patterns in the hand-parallelized codes with the Cetus implementation scheme. We measured the execution time of those code sections and the entire program.

Table~\ref{table:ArrayReduction} shows a significant performance impact on the two loops. The overall effect in the overall programs is minor, as loop rhs\_norm\#1 is small and executed once per application in both SP and BT. 
In general, as reduction operations can show up in compute-intensive sections of programs, the impact may be much larger, however.

\begin{table}[H]
 \caption{The table compares the performance of the Cetus-applied array-reduction transformation versus the manually applied technique in the hand-parallelized codes.}
\centering 
\begin{tabular}{| c c c c H c|} 
\hline\hline 

\multicolumn{1}{|p{3cm}|}{\centering Application Name}
& \multicolumn{1}{|p{2.5cm}|}{\centering Loop Name }
& \multicolumn{1}{|p{3.5cm}|}{\centering Technique Not Applied (Cetus)}
& \multicolumn{1}{|p{3cm}|}{\centering Technique Applied (Manual)}
& \multicolumn{1}{H}{\centering Performance Difference }
& \multicolumn{1}{|p{2cm}|}{\centering Impact}\\\hline
\hline 
 BT & rhs\_norm\#1  &0.005 & 0.003 &    0.676 & 66\%\\
 BT  &  \textbf{program}  & 117  &  116&    0.996 & 1\%\\
 SP  & rhs\_norm\#1 & 0.006  &0.002 &   0.333  & 200\%\\
 SP  &  \textbf{program} & 112  & 111 &    0.999 & 1\%\\[1ex] 
\hline 
\end{tabular}
\label{table:ArrayReduction} 
\end{table}

\subsubsection{\textbf{Opportunities for Compilers} }
Compilers can easily transform array reductions in either of the described variants. The choice of best implementation depends on several factors, including the efficiency of implementation of the OpenMP atomic directive. If the implementation simply uses a critical section (which we have seen in some OpenMP libraries), the current Cetus transformation likely performs the same or better.
This calls for compilers having knowledge of architectural and platform parameters, which we will discuss more in  Section~\ref{ConditionalParallelization} on conditional parallelization.

\subsection{\capitalisewords{NOWAIT -- Eliminating Barrier Synchronizations}}
\label{NOWAITPattern}
\subsubsection{\textbf{The Pattern} }A barrier is implicitly included at the end of some OpenMP constructs, including \textit{parallel, for, and single constructs}. 
This barrier is the safe default so that all threads have completed their share of work before proceeding with the execution. This synchronization is not needed if threads do not access data previously operated on by a different thread or on the last worksharing loop inside a parallel region. The OpenMP NOWAIT clause eliminates the barrier.

NOWAIT clauses have been inserted on many parallel loops inside the parallel regions in the hand-parallelized programs, reducing substantial overhead. The auto-parallelized code do not include such techniques.

\subsubsection{\textbf{Performance Impact} }
In order to test the performance impact of the technique, we have created program variants of hand-parallelized codes with removed NOWAIT clauses. Table~\ref{table:NOWAIT} on page \pageref{table:NOWAIT} compares these codes with the hand-parallelized variants. The impact in most of the programs is only about 1\%, even though individual loops see a gain of up to 16\%. It is likely that the impact would increase with a larger number of threads (recall that we use four) or in programs/loops with imbalanced load.

\subsubsection{\textbf{Opportunities for Compilers} }
Compile-time techniques for barrier elimination have been explored. Engineering them into available compilers is still an opportunity to be seized. Given the relatively minor impact, other techniques may need to be prioritized. Note also that this technique is related to enclosing loops in a parallel region.

\begin{table}[H]
\caption{Impact of Eliminating Barrier Synchronization: Execution times of removed versus present NOWAIT clauses in hand-parallelized codes. } 
\centering 
\begin{tabular}{| c c c c c H c|} 
\hline\hline 

 \multicolumn{1}{|p{2cm}|}{\centering Application Name}
& \multicolumn{1}{|p{1.8cm}|}{\centering Loop Name }
& \multicolumn{1}{|p{3cm}|}{\centering Number of NOWAIT Clauses }

& \multicolumn{1}{|p{2.5cm}|}{\centering Technique Not Applied}
& \multicolumn{1}{|p{2.5cm}|}{\centering Technique Applied }
& \multicolumn{1}{H}{\centering Performance Difference }
& \multicolumn{1}{|p{2cm}|}{\centering Impact}\\\hline
\hline 
 CG & Conj\_grad\#0-4 &1  & 0.173 & 0.149& 0.863 & 16\%\\
 CG &  \textbf{program}  &1 & 3 & 2.98 & 0.057  & \textbf{1}\%\\

  MG & norm2u3\#0 &1  &  0.121 & 0.119 & 0.987  & 2\%\\
  MG & comm3\#0  &1 & 0.003 &  0.003  & 1.008 & 0\%\\
  MG & \textbf{program}  &2 &  8.7 & 8.5 & 0.982  & \textbf{2}\%\\

  BT & initialize\#0-\#7 &6  & 0.116 & 0.116  & 0.998 & 0\%\\
  BT & exact\_rhs\#0-\#4  &2 & 0.142 & 0.142   &  0.995 & 0\%\\
  BT & compute\_rhs\#0-\#10 &7  &  21.343 & 20.345  &  0.998 & 5\%\\
  BT & error\_norm\#1  &1  & 0.019 &  0.019  &  0.999 & 0\%\\
  BT & rhs\_norm\#1  &1   & 0.003 &  0.003  & 0.989 & 0\%\\
  BT & \textbf{program}  &17  &  117 & 116  & 0.994  & \textbf{1}\%\\

  SP & initialize\#0-\#7  &6  & 0.043 &  0.043  &  1 & 0\%\\
  SP & exact\_rhs\#0-\#4 &2   &  0.133 & 0.132 &  0.001 & 1\%\\
  SP & compute\_rhs\#0-\#10 &7 &  21.452 & 20.050 & 1.401  & 7\%\\
  SP & error\_norm\#1 &1     & 0.012 & 0.011 &  0.001 & 9\%\\
  SP & rhs\_norm\#1  &1   & 0.002  &  0.002  & 0.0003  & 0\%\\
  SP & \textbf{program} &17  & 111.5 & 110.5  & 0.985  & \textbf{1}\%\\[0ex] 
\hline 
\end{tabular}
\label{table:NOWAIT} 
\end{table}

\subsection{\capitalisewords{Conditional Parallelization}}
\label{ConditionalParallelization}
\subsubsection{\textbf{The Pattern} }A technique present in auto-parallelized codes, but not in the manual variants, is the conditional parallelization of loops. The Cetus compiler estimates the workload of loops, as the product of the number of statements and iterations. It parallelizes only those with high workloads. If the estimate is an expression that cannot be computed at compile time, it uses OpenMP's conditional parallelization clause with an if condition that the expression exceeds a threshold. The manually parallelized programs do not use such conditional parallelization. One can expect that conditional parallelization benefits programs with small data sets, as some of the loops will not beneficially run in parallel.

\subsubsection{\textbf{Performance Impact} }
We have found some conditionally parallelized loops that are too small to run in parallel beneficially. But their impact on the overall programs was very small, even when they were executed in parallel, as these loops do not take significant execution time. We have also found some loops that Cetus did not parallelize since they were not profitable. While conditional parallelization adds a small runtime overhead, due to the if clause added to the OpenMP directive that checks for exceeding  a threshold, the technique is generally beneficial. 

To estimate the overhead conditional parallelization may add to the execution time of the loop, we measured the execution time of the rhs\_norm\#1 loop of the BT application
with and without conditional parallelization in the Cetus-parallelized code. The results of this measurement are presented in Table \ref{table:conditionalParallelization}.

\begin{table}[H]
\caption{Impact of conditional analysis; A comparison is made between the execution time of the code with and without conditional analysis in Cetus-parallelized code.
} 
\centering 
\begin{tabular}{| c c c c H c|} 
\hline\hline 

\multicolumn{1}{|p{3cm}|}{\centering Application Name}
& \multicolumn{1}{|p{2.5cm}|}{\centering Loop Name }
& \multicolumn{1}{|p{3cm}|}{\centering Technique Applied }
& \multicolumn{1}{|p{3cm}|}{\centering Technique Disabled }
& \multicolumn{1}{H}{\centering Performance Difference }
& \multicolumn{1}{|p{2cm}|}{\centering Impact}\\\hline
\hline 
 
 BT  & rhs\_norm\#1 &  0.004 &   0.003 & 0.001  & 33\%\\[1ex] 
\hline 
\end{tabular}
\label{table:conditionalParallelization} 
\end{table}
 \subsubsection{\textbf{Opportunities for Compilers} }
Like many other optimizers, Cetus uses a crude profitability model. There is much room for improving these models to estimate the execution time of loops or program sections under various optimization variants. 

\subsection{\capitalisewords{Code Modifications in hand-parallelized codes}}
\label{CodeModifications}

\subsubsection{\textbf{The Pattern} }Our comparison of hand-parallelized and auto-parallelized versions of the codes revealed additional modifications that were made to the hand-parallelized codes. They include:
\begin{itemize}
     \item Enclosing a mix of parallel loops and serial sections in a parallel region. An example of this is in the CG application, loops conj\_grad\#0-\#4.
     \item Changing the scope of variables to enable the creation of parallel regions. An example of this is in the CG application, conj\_grad() subroutine.
     \item Explicitly mapping tasks or iterations to threads. An example is the create\_seq\#0 loop in the IS application.
    \item Resolving some dependences (mostly output dependences) to enable parallelization. An example is the IS application's full\_verify\#0 loop.
    \item Improving cache performance. An example is the rank() subroutine in the IS application.
    \item Merging loops that perform similar tasks. An example is the MG application's comm3\#0 loop.
    \end{itemize}

\subsubsection{\textbf{Performance Impact} } These modifications were often applied to enable later parallelization steps. Their impact was thus difficult to isolate. In general, they contributed significantly to the good performance of the parallel applications.

\subsubsection{\textbf{Opportunities for Compilers} } While, at a high level, the mentioned modifications seem automatable, they tend to make use of application-specific knowledge. They are thus non-trivial to implement with general benefit.

\section{Related Work}
\label{relatedwork}
Many studies have been conducted on the NAS Parallel Benchmarks, including analyses, evaluation, parallelization, and tuning, but no comparison has been made between automatically and manually parallelized codes. Some studies do propose improvements to auto-parallelizers, based upon limitations of such tools that they have encountered in their experiments. The following are studies in this regard.

A study by Prema et al.\cite{Prema_Jehadeesan_Panigrahi_2017} comparing different auto-parallelizers such as Cetus, Par4all~\cite{Amini}, Pluto~\cite{10.1145/1375581.1375595}, Parallware~\cite{7228880}, ROSE~\cite{quinlan2011rose}, and Intel C++ Compiler (ICC)~\cite{tian2002intel} while parallelizing NAS Parallel Benchmarks (NPB) finds that auto-parallelizers have limitations, which programmers should be aware of
in order to intervene manually if necessary during parallelization. The development of an interactive environment that highlights the difficulties encountered while parallelizing loops is proposed.

Blume~\cite{Blume_1992} and Eigenmann et al.\cite{eigenmann1998automatic} discuss the successes and limitations of auto-parallelizers, based on a study performed on the Perfect Benchmarks. A modified version of the KAP restructurer and the VAST restructurer were used as representatives of parallelizing compilers for Fortran programs. Based on the limitations of auto-parallelizers at that time, this study proposes new techniques.

Dave et al.~\cite{Dave_Eigenmann_2010} have measured the serial performance as well as the performance of the manually parallelized codes on a subset of the NAS Parallel and SPEC OMP2001 benchmarks. In contrast to the present paper, their experiment compared the performance of these programs with that of auto-tuned codes.

A distinguishing feature of the present paper is that it compares auto-parallelized with hand-parallelized codes in order to identify opportunities for compiler improvements. Performance differences attributable to the identified program patterns are also measured, so as to quantify their importance for future compiler developments. Our proposed improvements could help auto-parallelizers reach performance approaching that of hand-parallelized code.

\section{Conclusion}
\label{conclusion}
We compared how expert programmers parallelize programs with how automatic parallelization does the same. The goal is to learn how to improve auto-parallelizers so as to approach hand-optimized performance. We believe that such studies are essential to push the forefront of research in compilers for parallel computing.

Currently, auto-parallelized codes are not as efficient as hand-parallelized codes. Our analysis of a subset of the NAS Parallel benchmarks found that auto-parallelized codes perform better than serial codes in many programs, but hand-parallelized codes perform significantly better. We have identified code sections, their program patterns, and the performance where differences occur. Additionally, we found examples in which hand-parallelized codes performed better after the use of threadprivate data was undone.

Among the patterns that showed the biggest performance differences were: Parallelizing the outermost loop in nested loops, parallelizing loops that enclose function calls, parallelizing loops with irregular patterns, enclosing loops in parallel regions, applying dynamic scheduling to cases with imbalanced loads, and using NOWAIT clauses to eliminate implicit barriers. 

Opportunities for advancing compilers exist in several areas, including in advancing analysis techniques, improving transforming techniques, and efficient OpenMP code generation. These opportunities are explained along with the differentiating patterns that have been identified.
The findings of this study will be used to guide future developments of the Cetus parallelizing compiler platform and the interactive Cetus parallelizer (iCetus)~\cite{Barakhshan_Eigenmann_2022}.

\subsubsection{Acknowledgements} This work was supported by the National Science Foundation (NSF) under Awards Nos. 1931339, 2209639, and 1833846.

%
%
%

\end{document}